\documentclass[copyright]{eptcs}
\providecommand{\event}{DCM 2012}

\usepackage{amssymb}
\usepackage{amsthm}

  \theoremstyle{remark}     \newtheorem*{aside}      {Aside}

  \newcommand{\qmod}[1]{{\textup{[}{#1}\textup{]}}}

\title{Ray tracing---computing the incomputable?}
\author{Ed Blakey
\institute{School of Mathematics, University of Bristol, University Walk, Bristol, BS8~1TW, United Kingdom}
\email{ed.blakey@queens.oxon.org}
}
\def\titlerunning{Ray tracing---computing the incomputable?}
\def\authorrunning{E.~Blakey}
\begin{document}
\maketitle

\begin{abstract}
We recall from previous work a model-independent framework of computational complexity theory.
Notably for the present paper, the framework allows formalization of the issues of precision that present themselves when one considers physical, error-prone (especially analogue rather than digital) computational systems.
We take as a case study the ray-tracing problem, a Turing-machine-incomputable problem that can, in apparent violation of the Church-Turing thesis, nonetheless be said to be solved by certain optical computers; however, we apply the framework of complexity theory so as to formalize the intuition that the purported super-Turing power of these computers in fact vanishes once precision is properly considered.

\end{abstract}

\section{Introduction}

There are certain unconventional models of computation that appear at first glance to offer super-Turing power or efficiency, only to prove impracticable when physically implemented.
Furthermore, the reasons for which the models transpire to be too good to be true are often difficult to formalize: it may be that any purported computational advantage, such as time/space complexity that compares favourably with that of the best known Turing machines, comes at a cost---such as an unrealistic constraint on the precision with which the computer must be constructed---for which there exists no complexity-theoretic formalism.
Rather, one must \emph{intuit} that a computational model is `cheating' (by deferring to unreasonable precision requirements the inherent difficulty of the computation, for example), and forming this intuition is an informal and ad~hoc process; whereas the comparatively formal and mechanical tools of complexity theory deal effectively with issues of \emph{time} and \emph{space}, they cater poorly if at all for non-standard resources such as \emph{precision}.

In the present paper, we recap from \cite{thesis} and related publications a model-independent framework of computational complexity theory, in which one may consider not only time and space but also non-standard resources such as precision, and not only Turing machines but also unconventional computers.
We also recall from \cite{reif} the ray-tracing problem, a Turing-undecidable problem that nonetheless appears decidable by certain optical systems.
We go on to consider the problem and the optical devices of \cite{reif} in the context of the model-independent complexity framework of \cite{thesis}, focussing in particular on issues of precision and the computability implications thereof; ultimately, we dispel any suggestion that the optical systems violate the Church-Turing thesis.

\subsection{Model-independent complexity theory}\label{sec:intmic}

We recall throughout Section \ref{sec:intmic} the model-independent approach to complexity theory advocated in \cite{thesis}.

\subsubsection{Complexity theory}

\emph{(Computational) complexity theory} is the study of computers' efficiency and of computational tasks' difficulty.
By quantifying the resources---\emph{run-time} and \emph{memory space}, for example---necessary for the successful execution of a task when using a given device, one measures the efficiency of this device; this bounds from above the efficiency of the \emph{optimal} device that performs the same task, and thus bounds from above the difficulty of the task itself.

See for example \cite{zoo,papa,sips} for much more detail concerning complexity theory.

\subsubsection{Turing machines}

In order that the concepts of complexity theory be well defined, it is necessary to specify which class of computational devices one is considering.
By far the most thoroughly studied model of computation is the \emph{Turing machine} \cite{turi}; accordingly, the resources traditionally considered in complexity theory are those---\emph{time} and \emph{space}---consumed by Turing machines.

The model's almost exclusive consideration is widely seen as unproblematic, due perhaps in no small part to the \emph{Church-Turing thesis} \cite{chur}, which suggests that the tasks computable by Turing machine are precisely those effectively computable by any finite algorithm, and due perhaps also to the thesis's \emph{extended version}, which suggests that not only the computability but also the complexity of computational tasks is in some suitable sense independent of the choice of computational model.

Hence, the theses suggest, one may choose with impunity an arbitrary computational model within which to work, and nonetheless gain general, model-independent insights.
However, we claim that there exist computational models that demand an approach to complexity theory different from that practised when considering Turing machines.
This approach is described in detail in \cite{thesis} and outlined in the remainder of Section \ref{sec:intmic}.

\subsubsection{Unconventional computers}

So as to make more concrete the preceding comments, we claim (and defer proof and further discussion to, for example, \cite{bJFac}) that there exists an analogue computer that factorizes natural numbers in polynomial time and space, but that nonetheless has an exponential resource overhead; specifically, the computer's `precision
complexity'---see Section \ref{sec:intmicpre}---is exponential.

Therefore, a complexity analysis following the pattern of standard complexity theory---explicitly, an analysis that heeds the Turing-machine resources of time and space, but no others---would reveal only the system's polynomial time and space requirements, whilst overlooking its true exponential complexity, which arises by virtue of requirements relating to neither time nor space, but rather to precision.

Note, then, that this analogue system is an example of an unconventional computer that requires in its analysis an unconventional approach to complexity theory.
We suggest that other non-standard computational paradigms---quantum, chemical and, notably for the purposes of the present paper, \emph{optical} systems, for example---may similarly demand non-standard complexity-theoretic techniques.

\subsubsection{Precision}\label{sec:intmicpre}

We hint above that, although our analogue factorization system \cite{bJFac} requires only polynomial quantities of time and space, it nonetheless has an exponential precision overhead.
Implicit, then, is that one may quantify precision as a computational resource, whence the corresponding complexity function may be considered; we briefly recall now the method whereby this quantification may be performed, though defer full details to, for example, \cite{bJFac,thesis}.

Let us be more specific about what one wishes to quantify: the aim is numerically to measure the precision required of a system's user during the system's input and output processes.
It may be, for example, that the user conveys his chosen input value to some physical computing system by manipulating a physical parameter of the system---e.g.,\ turning a dial through an angle, or adjusting the wavelength of a source of rays; the angle/wavelength encodes the input value\mbox{---,} and that the user receives the output value from the system by measuring a physical parameter.
If these input and output processes are subject to imprecision, as with physical computing systems they are indeed likely to be, then the user's intended input value and that actually received by the system may differ, as may the actual output value supplied by the system and that measured by the user.
Suppose now that these differences can be parameterized numerically: say that the intended and actual input values differ by no more than $\epsilon_1 \geq 0$, and the actual and measured output values by no more than $\epsilon_2 \geq 0$.
(We describe here \emph{additive} error, where imprecision corrupts a true value $x$ so as to render it an arbitrary element of $\left[x - \epsilon_i, x + \epsilon_i\right]$ ($\epsilon_i \geq 0$); one may equally consider other forms of imprecision, such as \emph{multiplicative} error, where $x$ is rendered an arbitrary element of $\left[\frac{x}{\epsilon_i}, \epsilon_i x\right]$ ($\epsilon_i \geq 1$). Which of these or other models of imprecision is appropriate in a given situation depends upon the details of implementation of the process whence the imprecision arises.)
Suppose further that sufficiently small error pairs $\left(\epsilon_1, \epsilon_2\right)$ are corrigible: that, provided that the $\epsilon_i$ are sufficiently close to $0$ (if the error is additive, or $1$ if multiplicative, etc.),\ the computation is successful in that the user receives the correct output value corresponding to his intended input value, imprecision notwithstanding.
Then it is natural to consider the \emph{area} of the region of corrigible pairs $\left(\epsilon_1, \epsilon_2\right)$ within the two-dimensional space with axes
$\epsilon_i$---see (\ref{eq:trireg}) below for an example of such a region; the larger the area, the more robust against input/output imprecision the system.
One takes one divided by this area as the quantification of required precision, and views the quantity as a function of the size of the input value so as to obtain a complexity function: \emph{precision complexity}.
(We give here only a feel of the quantification of precision, and reiterate that \cite{bJFac,thesis} offers much more detail and justification.)

\subsubsection{Resource}\label{sec:intmicres}

We note from the above discussion that the crucial shortcoming of a standard complexity analysis in the case of our analogue factorization system \cite{bJFac} is that such an analysis overlooks relevant \emph{resources}, in this instance precision.
Accordingly, a feature of our model-independent framework of complexity theory is that it accommodates \emph{arbitrary} resources: as well as the standard \emph{time} and \emph{space}, and the illustrative, non-standard \emph{precision} outlined above, one may consider other non-standard resources such as \emph{energy}, \emph{mass}, \emph{thermodynamic cost} and \emph{material cost}, for example.

For present purposes, a \emph{resource} $A_\Phi$ is a function that maps each valid input value $x$ that may be passed to a computing system $\Phi$ to the corresponding natural number $A_\Phi\left(x\right)$ of units of $A$ used by $\Phi$ given input value $x$ (this definition indeed suffices ``\qmod{f}or present purposes'', though it is beneficial in certain other contexts to restrict oneself to consideration only of so-called \emph{normal}
resources---see \cite{bJRes} for discussion of this side-issue).
The \emph{complexity function} $A^*_\Phi$ corresponding to resource $A_\Phi$ is defined by $A^*_\Phi\left(n\right) := \sup\left\{\, A_\Phi\left(x\right) \;\middle\vert\; \left\vert x\right\vert = n \,\right\}$, where $\left\vert x\right\vert$ denotes the \emph{size} of input value $x$.

For example, let $T_\Phi\left(x\right)$ be the number of \emph{time-steps} that elapse, and $S_\Phi\left(x\right)$ the number of \emph{tape-cells} written to, during the computation performed by Turing machine $\Phi$ given input value $x$; then $T_\Phi$ and $S_\Phi$ are the standard resources of \emph{time} and \emph{space}, and $T^*_\Phi$ and $S^*_\Phi$ the corresponding complexity functions.
More generally than this, however, the model-independent complexity framework allows $\Phi$ to be a computational device other than a Turing machine, and one may consider resources other than time and space.

See \cite{ucuc} for further discussion of resources, and, more fundamentally, of interpretations of the term `resource' itself.

\subsubsection{Dominance}

Having introduced into our framework arbitrary resources, we have solved the difficulty of overlooking relevant complexity measures such as precision, only to face a new problem: quantifying computers' \emph{overall} complexity.

In the Turing-machine case---in which the only resources are time and space, of which the former is always consumed in the greater quantity\mbox{---,} time complexity gives an adequate measure of overall complexity; hence, computers' relative efficiency may be ascertained by considering their respective \mbox{time- (i.e.,}\ overall-) complexity functions within the preordering `$\lesssim$', where `$f \lesssim g$' denotes `$f \in \mathcal{O}\left(g\right)$'.

However, in the case of general computation, where, in line with the discussion above, one may consider many different resources---time, space, precision, and so on\mbox{---,} it is no longer clear how to abstract a measure of overall complexity, whence to compare computers' efficiency.

We recall from \cite{bJRes,thesis}, to which we defer further detail, that, relative to a fixed, finite, non-empty set $\mathcal{R}$ of resources consumed by computer $\Phi$, a resource $A_\Phi \in \mathcal{R}$ is said to be \emph{dominant} for $\Phi$ if it is maximal in the preordering $\left(\mathcal{R}, \lesssim\right)$; i.e.,\ if $A_\Phi \lesssim B_\Phi \Rightarrow B_\Phi \lesssim A_\Phi$ for all $B_\Phi \in \mathcal{R}$.
Then we define the \emph{overall complexity} $\mathcal{B}^*_{\mathcal{R}, \Phi}$ of a computer $\Phi$, again relative to $\mathcal{R}$ and again with detail deferred to \cite{bJRes,thesis}, to be the sum of all complexity functions corresponding to dominant resources: $\mathcal{B}^*_{\mathcal{R}, \Phi}\left(n\right) := \sum_{A_\Phi\textrm{\scriptsize \ is\ }\mathcal{R}\textrm{\scriptsize -dominant}} A^*_\Phi\left(n\right)$.

\subsection{Ray-tracing problem}\label{sec:intray}

We describe now the situation to which we apply the framework recapped above.

Reference~\cite{reif} discusses a paradigm of \emph{optical computation} that sees the values with which computation is performed, including intermediate values analogous to a Turing machine's mid-computation tape contents, encoded as spatial coordinates of rays of light, and sees the operations applied to these values implemented using reflection/refraction by surfaces.
The \emph{ray-tracing problem}---see \cite{reif} for more detail---asks, given the set-up of an optical system conforming to this paradigm and given the initial position and direction of a ray of light, whether the ray ever reaches a given point.
The problem, which is, in a sense, naturally and efficiently solved by such an optical system, is nonetheless shown in \cite{reif} to be undecidable by Turing machine (further, \cite{reif} demonstrates that this undecidability may persist even if the objects forming the optical system are finitely expressible, e.g.,\ via rational quadratic/linear equations), which suggests of the abstract optical model super-Turing power.
(Reference~\cite{reif} focuses on issues of the \emph{Turing}-computability---or lack thereof---and \emph{Turing}-complexity of the ray-tracing problem, in particular without considering the non-Turing, \emph{optical} systems as a serious alternative, more powerful computing paradigm. Nonetheless, the findings of that paper prompt such consideration, which we accordingly make here.)

\subsection{Motivation}

Prompted by the above suggestion that, at least as mathematical abstractions if not as real-world implementations, certain optical computers boast super-Turing power, we formalize in Section \ref{sec:prt} certain issues, notably those pertaining to precision, surrounding the ray-tracing problem and the optical devices of \cite{reif}.

We note at this juncture in defence of the authors of \cite{reif} that they are clear from the outset that they deal with a theoretical model that is based upon the assumption that optics works in an idealized, geometric way, with no wave phenomena, etc.
For example, one reads in their description of the ray-tracing problem that
\begin{quote}``\qmod{r}ays are assumed to have infinitesimal wavelength and are treated as lines with zero width. This implies that there is no diffraction caused by the wave nature of light. All surfaces are perfectly smooth and do not cause the scattering of rays upon reflection or refraction''.
\end{quote}
Later, one reads of the computational paradigm under consideration that,
\begin{quote}``\qmod{t}heoretically, these optical systems can be viewed as general optical computing machines, if our constructions could be carried out with infinite precision, or perfect accuracy. However, these systems are not practical, since the above assumptions do not hold in the physical world. Specifically, since the wavelength of light is finite, the wave property of light, namely diffraction, makes the theory of geometrical optics fail at the wavelength level of distances''.
\end{quote}

This context justifies the lack in \cite{reif} of consideration of the precision issues inherent in these optical systems.
When, however, such issues \emph{are} considered---when the distinction is drawn between mathematical model and implemented computer\mbox{---,} one may expect that the problem actually being solved by the system, as impaired by its physically imposed limitations, is no longer (Turing-) undecidable; formalization within our complexity framework of the situation allows us to demonstrate in Section \ref{sec:prt} that this is indeed the case.

\section{Precision and the ray-tracing problem}\label{sec:prt}

We recap above the constituent concepts---resource, dominance, overall complexity, etc.---of our model-independent framework of complexity theory \cite{thesis}, and recall from \cite{reif} the ray-tracing problem and suggestions of its computability by optical systems, Turing-\emph{un}computability notwithstanding.
In this section, we apply the notions of the framework to the specific case study of these optical computers and their approach to the ray-tracing problem, so as to formalize the role of precision in this context.

\subsection{Precision complexity of storage/retrieval of a value}\label{sec:prtsto}

The ray-tracing methods described in \cite{reif} make use of encodings whereby entire tape contents are encoded as the angle\footnote{In fact, this angle is an intermediate encoding, later converted into an encoding as a spatial position. For the purposes of the present discussion, consideration of either encoding---angle or position---is adequate (notably, one encoding imposes an exponential precision requirement if and only if the other encoding does); we arbitrarily chose the former.} of a single ray of light, with one bit of precision in this angle being used to convey the contents of each tape cell to which is written.
(That the encodings are of tape contents implies that computations via these ray-tracing methods are instantiations of intermediate Turing-machine implementations; this imbues the optical model with universality in the Turing sense, but may increase the complexity of some problems---recall from \cite{step} Susan Stepney's thesis that it is more desirable to let computers function `naturally' than to coerce them into instantiating logic gates, or in this case Turing machines, or similar.)

We consider now the precision complexity of this process in the simple case where one merely \emph{stores} as an angle an $n$-bit value---which one may identify with $n$ cells' contents\mbox{---,} only to \emph{retrieve} it again.
Suppose that the system is two-dimensional, and, in particular, that the angle that encodes the value is a \emph{plane angle} in the range $\left[0, 2\pi\right)$, where the unit is the radian, and where we deem the interval to be `circular' in that $0$ and $2\pi$ are coincident, and, more generally, arithmetic is performed modulo $2\pi$.

The intention is that the $n$-bit input (i.e.,\ stored) value $x = \sum_{i = 1}^n 2^{-i} b_i \in \left[0, 1\right)$, where the $b_i \in \left\{0, 1\right\}$ are bits, be encoded as angle $\theta_x := 2\pi\left(x + 2^{-n - 1}\right)$---this is central in the range of angles $2\pi y$ such that $x$ and $y$ agree in the first $n$ bits after the binary point.
(In order to store and retrieve an arbitrary $n$-bit value, one needs sufficient precision to distinguish $2^n$ different states; the pigeon-hole principle dictates, then, that at fewest one such state is represented under the chosen encoding---whereby angles encode states---by a set of angles of measure at most $2\pi 2^{-n} = 2^{1 - n}\pi$. This bound is attained by our chosen scheme whereby each $n$-bit value $x \in \left[0, 1\right)$ is encoded as $\theta_x$, which is central in $\left[2\pi x, 2\pi \left(x + 2^{-n}\right)\right)$, an interval of measure $2^{1 - n} \pi$ disjoint with the analogous intervals corresponding to $n$-bit values other than $x$.)

Suppose now that the process is hampered by an additive input error of $\epsilon_1$: an intended input value $x \in \left[0, 1\right)$ is encoded as an arbitrary implemented angle $\theta_x' \in \left[\theta_x - \epsilon_1, \theta_x + \epsilon_1\right)$.
Suppose, similarly, that there is an additive output error of $\epsilon_2$: angle $\theta_x'$ is measured as an arbitrary angle in the interval $\left[\theta_x' - \epsilon_2, \theta_x' + \epsilon_2\right)$.
Measurement of an encoding of $x$, then, will yield an arbitrary value $\theta'' \in \left[\theta_x - \epsilon, \theta_x + \epsilon\right)$, where $\epsilon = \epsilon_1 + \epsilon_2$; this is decoded via $\theta'' \mapsto \frac{\theta''}{2\pi} - 2^{-n - 1}$ as $x'' \in \left[x - \frac{\epsilon}{2\pi}, x + \frac{\epsilon}{2\pi}\right)$.

In order that true value $x$ and stored, retrieved value $x''$ agree to $n$ bits---and given that $x$ is encoded centrally in its range, as described above\mbox{---,} it is necessary that $x'' \in \left[x - 2^{-n - 1}, x + 2^{-n - 1}\right)$, which entails that $\frac{\epsilon}{2\pi} \leq 2^{-n - 1}$, i.e.,\ that $\epsilon \leq 2^{-n}\pi$ (for then the interval $\left[x - \frac{\epsilon}{2\pi}, x + \frac{\epsilon}{2\pi}\right)$ in which $x''$ is known to lie is a subset of the interval $\left[x - 2^{-n - 1}, x + 2^{-n - 1}\right)$ of values in agreement to $n$ bits with $x$).
Hence, the region of corrigible error pairs $\left(\epsilon_1, \epsilon_2\right)$---recall Section \ref{sec:intmicpre}---is
\begin{equation}\label{eq:trireg}
\left\{\, \left(\epsilon_1, \epsilon_2\right) \in \mathbb{R}^2 \;\middle\vert\; \epsilon_1, \epsilon_2 \geq 0 \wedge \epsilon_1 + \epsilon_2 \leq 2^{-n}\pi \,\right\} \enspace ,
\end{equation}
a triangle of measure $2^{-2n - 1}\pi^2$, whence the required precision of this store-and-retrieve system $\Phi$ given $n$-bit input $x$ is $P_\Phi\left(x\right) = 2^{2n + 1}\pi^{-2}$.

Thus, the precision complexity $P^*_\Phi$ of the process of merely storing and retrieving a value is \emph{exponential} in the number $n$ of tape cells written to by an intermediate Turing-machine implementation: $P^*_\Phi\left(n\right) = 2^{2n + 1}\pi^{-2}$.

\subsection{The `finite-precision ray-tracing problem' is \emph{not} incomputable}\label{sec:prtfin}

This increasing, let alone exponentially increasing, precision complexity has the following consequence: if the precision available to a user of such a system as described in \cite{reif} is a~priori bounded, e.g.,\ by technological factors---which, in practice, it necessarily will be\mbox{---,} then the size, too, of problem instances that can be processed by the system is bounded.
The system solves, therefore, not the undecidable ray-tracing problem in its entirety, but rather a proper subproblem, admitting only bounded input instances and concerning only physically realizable optical configurations rather than those with coordinates drawn arbitrarily from the continuum of real numbers or even from the countable infinity of rational numbers.
This subproblem, we claim, is \emph{computable by Turing machine}; this is because the effective discretization of positions, angles, etc.\ of an optical system's constituent parts brought about by the imposition of fixed, finite precision renders merely countably many the distinguishable such systems.
The corresponding ray-tracing problem restricted to these countably many systems is then computable by Turing machine, e.g.,\ by simulation of the optical system for only as long as is required to achieve sufficient precision.
Thus, the findings of \cite{reif} are not in fact in conflict with the Church-Turing thesis.

\begin{aside}
A more striking and sadly lacking resolution of the apparent conflict between \cite{reif} and the Church-Turing thesis would see required precision shown to be \emph{infinite}, not merely in the limit, but also for some finite input size; this is simply not the case, however (at least, not for the `commodity' resource of precision---the `manufacturing' resource is a different matter;
see Section \ref{sec:prtinf}).
Whereas an exponentially increasing precision is required to implement the methods described in \cite{reif}, it is at least a \emph{finite} precision---the method is precluded not \emph{in principle}, but only \emph{technologically} (say), by issues concerning precision.
We recall in particular that \cite{reif} discusses the computation of the \emph{incomputable} via such methods; one might, given the Church-Turing thesis, therefore expect preclusion \emph{in principle}, which, as explained, is not the case.
\end{aside}

The above aside notwithstanding, however, we reiterate that, concretely and technologically, precision bounds the system's processable input sizes; additionally we note that, abstractly and fundamentally, precision---and, in particular, its increasing---renders problematic certain seemingly innocuous assumptions (that diffraction is not present, etc.)\ when input values become sufficiently large.
The idealized model of optical computation does indeed seem capable, for what it is worth, of computing the incomputable, but real-life implementations necessarily suffer because of the supposedly subtle differences, which become less subtle for large input, between the theoretical and practical models---differences that arise due to apparently harmless assumptions that quickly become problematic in the presence of exponential precision complexity.
The Church-Turing thesis is not, we suggest, violated: the proposed optical paradigm is, by virtue of unrealistic assumptions regarding precision, not a valid, real-world computational model.

This is brought into even starker focus upon consideration of the precision requirements when \emph{manufacturing} the optical systems, as we now explain.

\subsection{Infinite manufacturing precision}\label{sec:prtinf}

In line with Section \ref{sec:intmicres}, we consider in Sects.~\ref{sec:prtsto} and \ref{sec:prtfin} the role of precision as a `commodity' resource \cite{ucuc}, i.e.,\ as being required in some quantity \emph{during computation}.
However, the systems of \cite{reif} also incur a significant `manufacturing' precision cost, a precision requirement not at run-time but rather during the \emph{construction} of the apparatus;
see \cite{ucuc} for detailed discussion of the notions of commodity and manufacturing resource.

For example, Theorem~5.7 of \cite{reif}---which is not a direct prerequisite of, but certainly presents issues similar to those encountered by, optical systems that solve the ray-tracing problem---relies on the introduction to an optical system of \emph{irrational} numbers, which introduction is achieved via precise construction of lenses/mirrors with specific, irrational focal lengths, a rational approximation of which, \emph{however accurate}, not having the desired effect.
Explicitly, the physical implementation of such systems incurs an \emph{infinite} manufacturing precision cost (regardless of the size of the input value with which one computes).
Although the analysis above of \emph{commodity} resources fails to capture the impossibility of such systems' use, we have more fundamentally that \emph{manufacturing} resources preclude the construction to sufficient precision of these devices.
This suggests that the computational difficulty giving rise to the ray-tracing problem's Turing-incomputability is in fact deferred from the optical devices' run-time to their manufacturing stage; this gives the devices, \emph{once they have been brought into existence}, the appearance of having super-Turing power, but the still-present computational difficulty manifests itself instead in the impossibility of constructing these systems to sufficient precision in the first place.
Thus, the systems, unimplementable as they are, do not violate the Church-Turing thesis.

We emphasize that it is not the aim of the authors of \cite{reif} to assess \emph{practically implemented instances} of their systems; by their own admission, the authors are working relative to explicitly stated assumptions, and their findings are, in that context, perfectly correct.
What we formalize here is that these assumptions, by virtue of their precision implications, render the considered systems unsuitable as real-world computers, and consequently render the findings of \cite{reif} irrelevant to the question of whether hypercomputation is \emph{possible} in any real sense.
In particular, the findings do not contravene the Church-Turing thesis.

\section{Conclusion}

We see above that the apparently super-Turing optical systems of \cite{reif} are in fact plagued by (a)~exponentially increasing commodity precision complexity and (b)~\emph{infinite} manufacturing precision complexity.

Even in the absence of (b), (a) alone would have as a consequence an upper bound on the size of input values that the systems could successfully process---recall Section \ref{sec:prtfin}.
Given such a practical limit on commodity precision (this limit may, for example, be imposed by technological constraints, or by required precision's becoming sufficiently great that effects---wave phenomena, for example---`abstracted out' of a physical situation during its mathematical modelling cease to be negligible; we compare this comment with the observation of \cite{gandy} that Newtonian physics admits idealized systems that apparently violate the Church-Turing thesis, but that the unrealistic simplifications implicitly made when adopting Newtonian physics are \emph{necessary} for such violation), the problem actually solved by the system ceases to be undecidable, and the controversy, for practically implemented systems, is dispelled---only from a theoretical and abstract point of view do the systems demonstrate, for what it is worth, that the optical model, including unrealizable and often tacit assumptions about available precision, is strictly more powerful than the Turing machine.
(In fact, in order for such an optical system to achieve super-Turing power, one must assume not only unimplementable degrees of commodity and manufacturing precision, but also infinitely quickly propagating rays: the sense in which the device `solves' the ray-tracing problem is that one \emph{observes} whether or not the ray arrives at the destination point under consideration; even granted perfectly precise construction and functioning of the system, however, this observation process is, due to the finiteness of the speed of light, no more guaranteed to terminate than is a Turing-machine simulation of the path of the ray. Thus further dwindles the device's suitability for implementation.)

The preceding paragraph aside, (b) encapsulates the true difficulty---in fact, impossibility---of the ray-tracing problem as tackled by the optical systems of \cite{reif}.
The devices, however powerful they may be in principle, cannot be physically realized with sufficient precision to function as claimed.
The Church-Turing thesis says nothing of one's ability to posit \emph{impracticable} super-Turing systems, and in particular those of \cite{reif} do not contravene the thesis.

\subsection{Extent of resolution}

As a final note, we consider the extent to which the controversial claim of optical computers' super-Turing power is resolved by the discussion in the present paper.

We outline above a framework in which certain of the relevant issues may be, and above are, formalized; this enables pertinent questions---ultimately, we hope, `is the claim of super-Turing power true or not?'---to be formulated more precisely and answered more confidently.
However, it must be noted that the question actually answered is, `is the claim true or not \emph{according to the framework and the assumptions implicit in its definition}?';\ it is trivial, moreover, to posit some framework that merely \emph{deems} the claim to be valid, or to be invalid.

The important factor, then, is the success with which our \emph{formal framework} of resources, complexity, dominance, etc.\ captures one's \emph{intuitive understanding} of computational resource and related notions.
The question of the extent of this success is not susceptible to rigorous proof, being, much like the Church-Turing thesis, a question of equivalence between a formal and an intuitive notion; rather, the equivalence must be borne out by evidently sensible choice, and sustained successful use, of the framework's definitions---or else the equivalence fails.
The former test has hopefully been shown, in the discussion and justification of our definitions---see \cite{thesis} for more---, to have been passed (for now{\ldots});\ the latter is necessarily ongoing, with the present paper representing a step in the right direction.

\subsection*{Acknowledgements}

We thank Bob Coecke and Jo\"{e}l Ouaknine for their supervision of the project from which this work stems, as well as conference participants, collaborators and colleagues---and the associated thesis's examiners Peter Jeavons and John Tucker---who have helped to shape and direct this research.
We thank the anonymous DCM~2012 referees for their detailed and useful comments, and participants of that conference for their encouraging feedback and interesting discussion.
We acknowledge the generous financial support of both the EPSRC (this work forms part of grant EP/G003017/1) and the Leverhulme Trust (which funds the author's current position).

\nocite{*}
\bibliographystyle{eptcs}
\bibliography{blakey}

\end{document}